\documentclass[10pt,conference]{IEEEtran}
\IEEEoverridecommandlockouts
\usepackage{cite}
\usepackage{amsmath,amssymb,amsfonts}
\usepackage{algorithmicx}
\usepackage{algorithm}
\usepackage{algpseudocode}
\usepackage{graphicx}
\usepackage{textcomp}
\usepackage{xcolor}
\usepackage{ntheorem}
\usepackage{caption}
\usepackage{tikz}
\usetikzlibrary{shapes,decorations,arrows,calc,arrows.meta,fit,positioning}
\tikzset{
    -Latex,auto,node distance =1 cm and 1 cm,semithick,
    state/.style ={ellipse, draw, minimum width = 0.7 cm},
    point/.style = {circle, draw, inner sep=0.04cm,fill,node contents={}},
    bidirected/.style={Latex-Latex,dashed},
    el/.style = {inner sep=2pt, align=left, sloped}
}
\usepackage{subcaption}
\usepackage{comment}
\usepackage{hyperref}
\usepackage[textsize=small]{todonotes}

\DeclareRobustCommand*{\IEEEauthorrefmark}[1]{%
  \raisebox{0pt}[0pt][0pt]{\textsuperscript{\footnotesize #1}}%
}
\usepackage{url}
\def\BibTeX{{\rm B\kern-.05em{\sc i\kern-.025em b}\kern-.08em
    T\kern-.1667em\lower.7ex\hbox{E}\kern-.125emX}}

\begin{document}

\title{
Dataflow graphs as complete causal graphs
\thanks{Correspondence to: ap2169@cam.ac.uk}
\thanks{* Equal contribution}
}

\author{
  \IEEEauthorblockN{%
    Andrei Paleyes*\IEEEauthorrefmark{1},
    Siyuan Guo*\IEEEauthorrefmark{1}\IEEEauthorrefmark{2},
    Bernhard Schölkopf\IEEEauthorrefmark{2},
    Neil D. Lawrence\IEEEauthorrefmark{1}
  }
  \IEEEauthorblockA{\IEEEauthorrefmark{1}\textit{Department of Computer Science and Technology, University of Cambridge} \\
 \IEEEauthorrefmark{2}\textit{Max Planck Institute for Intelligent Systems}
  }
}


\newtheorem*{definition}{Definition}
\newtheorem*{example}{Example}
\maketitle

\thispagestyle{plain}
\pagestyle{plain}

\begin{abstract}
Component-based development is one of the core principles behind modern software engineering practices. Understanding of causal relationships between components of a software system can yield significant benefits to developers. Yet modern software design approaches make it difficult to track and discover such relationships at system scale, which leads to growing intellectual debt. In this paper we consider an alternative approach to software design, flow-based programming (FBP), and draw the attention of the community to the connection between dataflow graphs produced by FBP and structural causal models. With expository examples we show how this connection can be leveraged to improve day-to-day tasks in software projects, including fault localisation, business analysis and experimentation.
\end{abstract}


\section{Introduction}
Modern software engineering industry becomes increasingly data centric. There are two primary motivations behind this trend. First, businesses and research groups seek to deploy more data driven software solutions, including those powered by machine learning (ML), which means software engineers are more often faced with data availability and analysis requirements \cite{lewis2021software,paleyes2022challenges}. Second, the rise of DevOps as a discipline of software maintenance practices emphasises attention to metrics that help assess and investigate hardware and software health. These metrics build on a growing amount of technical data which, while not directly related to a business domain, still needs to be collected, stored and analysed \cite{forsgren2018devops}.

However, raw data is often not enough to explain certain behaviour or answer questions about a software system. Modern software applications are complex and consist of a large number of components. It is often necessary to understand causal relationships between these components in order to effectively use the data available to answer business and technical questions, as well as insure quality \cite{Clark2022TestingCI}. Absence of such understanding can harm developers' ability to maintain software in the long term, an effect known as ``intellectual debt'' \cite{zittrain2019intellectual}. Dataflow graphs of all data states and transformations can assist in understanding system's inner mechanics. Unfortunately most common software paradigms, object and service orientation, focus on protecting data, and make recovery of dataflow graphs a complex task \cite{Paleyes2022FBP,lin2013scaling,stopford2016data}.

In the present paper, we draw attention to dataflow design paradigms, specifically flow-based programming (FBP, \cite{morrison2010flow}), and their connection to structural causal models. In particular, we argue that dataflow graphs of software systems can be treated as complete causal graphs, and unlike other modern design methodologies dataflow design provides such graphs natively. Through a range of illustrative examples we articulate how the connection between dataflow graphs and causal models can aid troubleshooting, business analysis, and experimentation in software systems. 

We highlight the importance of causally-aware decision making. It can evaluate the effects of our actions and guide future policies. However, a key bottleneck of causal reasoning methods is their reliance on a causal graph for inference tasks. This limits the applicability of causal inference \cite{Dawid2010BewareOT}, and hence much of the theoretical work focuses on discovering causal structure from observational data \cite{Janzing2017, Guo2022}. We argue that FBP has the advantage in offering a known causal graph. We will discuss in Section \ref{section:causality} and \ref{sec:dataflow_graph_as_causal_graph} how it provides a compact and efficient representation to infer effects and attribute explanations to address practical engineering questions.

\section{Motivating examples}
\label{section:motivating-examples}
This section introduces several examples of typical problems we aim to address. The setting for this section is a fictional online coffee retailer CoffeeFlow. CoffeeFlow operates a supply chain of coffee beans: they order coffee from the vendors, store it in the warehouses, and deliver it to the customers worldwide. To automate routine tasks CoffeeFlow invested into the development of a software platform that models the entire supply chain process. Following modern practices of distributed system design, the platform is implemented with microservices. Each service is maintained by a separate team of software engineers.

\textbf{Example 1}. Nadia is a software developer on pager duty for a service that handles customer orders. She receives a notification that her service's latency is spiking. Nadia investigates and finds out that the latency increase is associated with a corresponding increase in the upstream service. She engages her colleague, who again identifies the problem up the call stack. Eventually the issue is located three levels up from Nadia's service, and is due to a memory shortage of the warehouse storage management service. The pager alarm should have gone to a different person, saving precious time. Can this troubleshooting experience be expedited? 

\textbf{Example 2}. Art is a business analyst at CoffeeFlow. He has recently noticed that company's sales in a certain region have dropped by 5\%, and investigates the issue. Art spends multiple days reviewing and analysing every business metric in that region and identifies several potential causes: some vendors delayed produce delivery, demand has diverged from the forecast over the past month, temporary shortage of staff at one of the warehouses. Art proceeds to produce a report which gives the best estimates of how likely, in his opinion, each issue affected the sales figure. Is there a way to make this analysis process more precise?

\textbf{Example 3}. Alice is an applied ML scientist. She has recently finished training a new forecasting model to improve customer demand prediction. Model quality metrics show considerable improvement. However Alice has concerns that some of the downstream services in the platform might have adapted to the behaviour of the previous forecasting algorithm, and thus model accuracy improvement might not translate into business value after the deployment. Can Alice validate this without running expensive experiments?

In subsequent sections we show how the connection between FBP and causality can address the questions raised here.

\section{Flow-based programming}
\label{section:fbp}
\begin{figure*}[t]
	\centering
	\includegraphics[width=\textwidth, height=4cm]{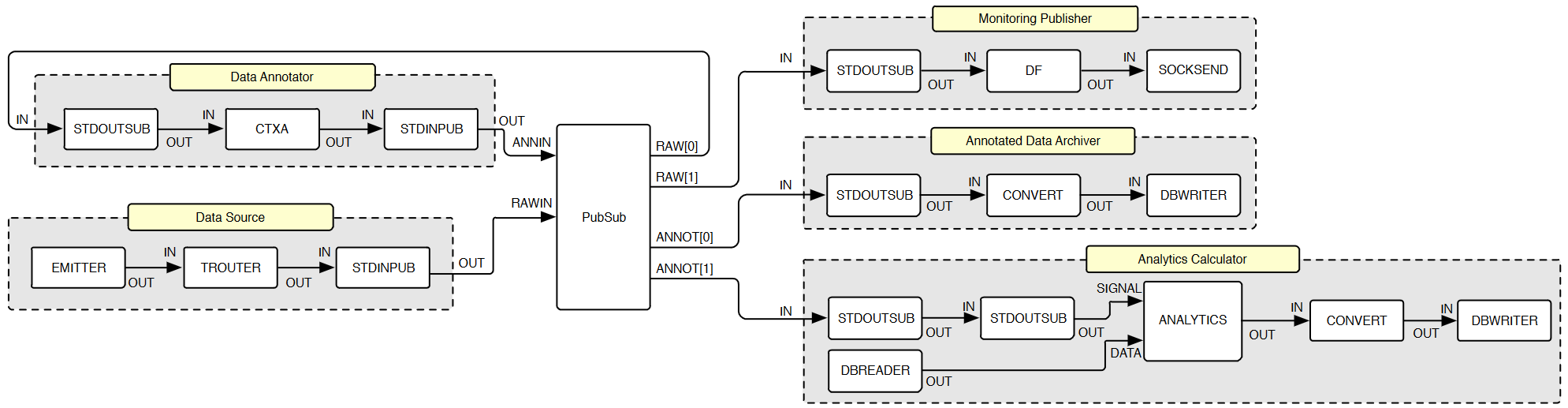}
	\caption{Example of a dataflow diagram of the FBP application for smart environments \cite{lobunets2014applying}. Crucially, this graph is created as an integral part of the software design process.}
	\label{figure:fbp_flow_diagram}
\end{figure*}

Flow-based programming (FBP) was introduced by J.P. Morrison in 1970s for distributed processing \cite{morrison2010flow}, and is considered a flavour of a more general dataflow programming paradigm, thus following the principles of dataflow computing architecture \cite{dennis75dataflow}. FBP defines a software system as a set of isolated processes that pass data between each other via connections that are external to those processes. Each process exposes data interfaces, known as named ports, that define inputs and outputs for that process. One of the key features of FBP-driven system design is that by defining data processing components and connections between them developers naturally build a dataflow graph of the entire software system, such as the one shown on Figure~\ref{figure:fbp_flow_diagram}.

Recently FBP started gaining momentum in the engineering community, particularly for development of ML workflows \cite{Mahapatra2022FlowBasedPF, akoush2022desiderata} and IoT pipelines \cite{szydlo2017flow}. Nevertheless FBP is still considered relatively niche \cite{sibirov2022fbpstate}. By highlighting the connection between FBP and causality we hope to spark the community's interest in this paradigm.

\section{Causality}
\label{section:causality}
We believe that the theory of causality can aid answers to the questions asked in Section~\ref{section:motivating-examples}. Here we give a quick introduction into its most relevant concepts.

Causality aims to discover causal relationships from observational data. Current statistical methods, in contrast, exploit associative relationships to improve prediction accuracy, which is sound as long as distributions do not shift (e.g., if the data is independent and identically distributed). But interventions in a system generally lead to distribution shift, and correlation does not imply causation. For example, the website {\em Spurious Correlation} \footnote{http://www.tylervigen.com/spurious-correlations} illustrates that the total revenue generated by arcade games is correlated with the number of computer science doctorates awarded in the US. Although this information can help us predict arcade revenue given the number of computer science doctorates awarded, it does not mean that if we want to improve the revenue of arcade we should award more doctorate degrees. The key difference is that rather than focusing on prediction, causality allows for actions. Below we introduce the language of structural causal model and show its natural connection with FBP.

\begin{definition}[Structural Causal Model (SCM) \cite{Pearl2009}]
A SCM $\mathcal{M}$ consists a set of variables $X_1, \dots, X_d$ and corresponding structural assignments of the form 
\begin{equation}
    X_i := f_i(PA_i^\mathcal{G}, U_i)
\end{equation}
for all $i \in \{1, \dots, d\}$ where $PA_i^\mathcal{G}$ are \textbf{parents} of $X_i$ in graph $\mathcal{G}$, referred to as \textbf{direct causes} of $X_i$. $U_1, \dots, U_d$ are noise variables, which we assume to be jointly independent. Here joint independence means for a sequence of random variables $U_1, \dots, U_d$, the joint distribution can be factorised as $p(u_1, \dots, u_d) = \prod_{i=1}^d p_i(u_i)$.
Further given a SCM $\mathcal{M}$, there is a corresponding directed acyclic graph (DAG) $\mathcal{G}$ where each variable $X_i$ has incoming edges from all the parent in $PA_i^\mathcal{G}$ to $X_i$.  
Any distribution generated by a SCM $\mathcal{M}$ can be factorised into causal conditionals via the \textbf{Markov factorisation}
\begin{equation}
    P(x_1, \dots, x_d) = \prod_i \underbrace{p(x_i | pa_i^\mathcal{G})}_{\text{causal conditional}}.
\end{equation}
\end{definition}

\begin{definition}[Intervention \cite{Didelez2012}]
An intervention is the act of affecting the system to control the outcome. In SCM it is denoted using the {\em do-operator}. Let $\sigma_X$ represent the intervention on random variable $X$. 
\begin{itemize}
\item Atomic intervention $\sigma_X = do(X=x_0)$: It sets the discrete random variable $X$ to a fixed value $x_0$ such that $p(x|pa_x^\mathcal{G}; \sigma_X) = \delta(x, x_0)$, where $\delta(x, y)$ takes value one if $x=y$ else it is zero.
\item Soft intervention: Instead of fixing value, we may want it to take values according to some user-defined distribution $q$ possibly depending on $PA_i^\mathcal{G}$:
$
p(x|pa_X^\mathcal{G}, \sigma_X = d_{pa_X}) = q(x|pa_X^\mathcal{G}).
$
\end{itemize}
\end{definition}

We can consider the intervention as functionality change in computational node and we would like to find out the effects of intervention without running the whole system. One natural solution is via \textbf{truncated factorisation} \cite{Pearl2009}. It transforms the post-intervention distribution to:
\begin{itemize}
    \item $P(x_1, \dots, x_d|\sigma_{X_i}) = 
    \prod_{j \neq i}p(x_j|pa_j^\mathcal{G}) \text{if} \ x_i = x_0 \ \text{else} \  0 $ under atomic intervention $\sigma_{X_i} = do(X_i= x_0)$. 
    
    \item $P(x_1, \dots, x_d|\sigma_{X_i}) =  q(x_i|pa_i^\mathcal{G}) \prod_{j \neq i}p(x_j|pa_j^\mathcal{G})$ under soft intervention $\sigma_{X_i} = d_{pa_{X_i}}$.
\end{itemize}

\section{Dataflow graph as causal graph}
\label{sec:dataflow_graph_as_causal_graph}
Object oriented and service oriented architectures, while being de-facto standard approaches in modern software engineering, fail to provide developers with complete data dependency graph of a system \cite{lin2013scaling,Paleyes2022FBP,nikolov2018oop,sharvit2022data}. Developers either have to use sophisticated and often commercial tools (e.g. Dynatrace \cite{dynatrace}, AWS X-Ray \cite{aws-x-ray}, Jaeger \cite{jaeger}), or invent custom ways to build data dependency graphs \cite{Baah2011MitigatingTC}. These approaches rely on strong assumptions about the system being analysed, and cannot guarantee completeness \cite{wang2021detecting, chowdhury2023evaluation}.

The key observation we put forth is that the dataflow graph produced in the course of building a software system with FBP is a complete data dependency graph of the entire system. For any given node all its upstream and downstream nodes can be found by traversing the graph. The design process guarantees that the graph is complete, with no hidden inputs or connections. Out-of-the-box availability and completeness mean this graph can be used immediately for causal inference on system components. FBP programs require no additional tools to enable causal reasoning and no assumptions about hidden confounders. In a recent review of applications of causality to software engineering \cite{siebert2022applications} it is noted that discovering the full data dependency structure of a system remains the main challenge for such applications. Dataflow architecture addresses this challenge, since the dataflow graph becomes available to developers as a direct output of the  design process.

We now introduce a causal attribution for flow-based programming. To our best knowledge, this is a first attempt to directly model software modules as causal conditionals. Often theoretical work focuses on attribution in graphical models without connection to software systems \cite{Singal2021Flow} and application oriented work requires extra effort to build data dependency graphs \cite{Baah2011MitigatingTC}. Algorithm \ref{alg:attribution_change} demonstrates how complete data dependency graph can help automate analysis and troubleshooting. Suppose we have $n$ data streams $\{X_i\}_{i = 1}^{n}$ and $m$ computational nodes $\{C_j\}_{j=1}^m$. For example, the circle nodes in Figure \ref{figure:insurance_claims} represent $C_j$ and rectangles represent $X_i$. Let $(C_j^{i}, C_j^{o})$ denote the set of (input, output) streams for a node $C_j$. Each record arrives in a data stream at time $t$, which is denoted as $X_{i, t}$. We assume $X_{i, t} \overset{i\,.i\,.d}{\sim} p(X_i)$ for all $t \in T$ within a user-defined time interval $T$. Note there are two kinds of reasons a change in the output stream can manifest: due to the change in an input data stream(s) and due to the change in a computational process(es).

\newlength{\textfloatsepsave}
\setlength{\textfloatsepsave}{\textfloatsep}
\setlength{\textfloatsep}{0pt}

\begin{algorithm}
\caption{Change attribution in software dataflow graphs}\label{alg:attribution_change}
\begin{algorithmic}[1]
\Require Dataflow graph with data streams $X_i$ and computational nodes $C_j$ ; $p \in \mathcal{D}^{new}, q \in \mathcal{D}^{old}$ distributions of all data streams in time frames $T^{new}, T^{old}$ respectively; target data stream $Y := X_d$
\State Model each data stream $X_i$ as a random variable.
\State Model each computational node $C_j$ with (input, output) data streams $(C_j^i, C_j^o)$ as conditional distributions. 
\State Initialise a dictionary of deviations $F_{dev}$ with computational nodes as keys.
\State Initialise a dictionary of attribution scores $F_{attr}$ with data stream nodes as keys.
\State Initialise an empty queue of data stream nodes $Q$.
\State Suppose we observe a change in $Y$ measured as $\Delta_Y = D(p(y)||q(y))$, where $D$ is KL divergence.
\State Push $Y$ to $Q$. 
\While{$Q$ is not empty}
\State Pop top element of $Q$ into a variable $S$
\State Find computational nodes $C_j$ that contribute to $S$.
\State Compute deviation of $C_j$'s functionality: \newline
\hspace*{1.2em} $\Delta_{C_j} = \textbf{Deviation}(p(s|pa_s), q(s|pa_s))$
\State $F_{dev}[C_j] = \Delta_{C_j}$
\For{each input data stream $ X_i \in C_j$}
\State Compute its contribution to the change in outcome: \newline
\hspace*{2.5em} $F_{attr}[X_i] = \textbf{Attribution}(X_i, \Delta_{C_j},  Y)$
\State Push $X_i$ to $Q$
\EndFor
\EndWhile
\State $\textbf{Aggregation}(F_{attr})$
\State \Return $F_{dev}$, $F_{attr}$

\end{algorithmic}
\end{algorithm}

The Algorithm \ref{alg:attribution_change} contains several subroutines, implementation of which can vary. \textbf{Deviation} for a computational node can be computed with KL divergence \cite{kullback1951information} or testing independence \cite{Budhathoki2021Why}. \textbf{Attribution} that computes scores reflecting how much a data stream node contributes to the observed shift in the output $Y$ can be calculated with Shapley values \cite{hart1989shapley} or proportional change of the KL divergence between the input data streams and output data stream, i.e. $\frac{\text{KL}(p(\text{input}) || q(\text{input}))}{\text{KL}(p(\text{output}) || q(\text{output}))}$. When proportional change is used, \textbf{Aggregation} backtracks multiple attribution values along the path from a given node to the output, in other cases \textbf{Aggregation} is a no-op.

We can now discuss how these observations and the techniques can allow practitioners to answer questions posed in section \ref{section:motivating-examples}, with an assumption that the CoffeeFlow platform was designed with dataflow paradigm instead of microservices.

\subsection{Fault localisation}
With the proliferation of ML, the engineering community started to develop ML-based methods for root cause analysis (RCA) and fault localisation in software \cite{ascari2009exploring, Wong2016ASO, zheng2016fault}. However, ML algorithms struggle to generalise well to rare events \cite{Lal2017RootCA}. Bugs that take significant developer time are not typical, and therefore time saving enabled by classical supervised and semi-supervised ML techniques is limited especially if automation is desired. However, the ability to automatically explain the origin of a rare event in software may arise from knowledge of causal relationships between system components. For example, Baah et al.~\cite{Baah2011MitigatingTC} analyse a computer program and produce two graphs: a data dependencies graph and a control dependencies graph. These graphs are then used as causal graphs to facilitate RCA. While this approach works well for a single program, their graph building method does not scale to complex systems consisting of multiple programs and services exchanging API calls.

Causal fault localisation in FBP programs can offer a better experience. A readily available dataflow graph can be used as an SCM with every graph node assigned a causal conditional distribution $p(x_i|pa_i^\mathcal{G})$, where $x_i$ is its output data stream and $pa_i^\mathcal{G}$ is its input data stream(s). Such conditional distribution contains the information of graph node's functionality and can be estimated directly via observing its (input, output) pairs. Change contribution can then be calculated according to Algorithm~\ref{alg:attribution_change}, and the largest deviation from independence will signify most likely faulty nodes. Moreover, relative magnitude of the contributions can be used to quantify uncertainty of the algorithm. In our example 1, Nadia was paged into a service latency issue that was caused by an upstream service. The troubleshooting process required engaging additional engineers and manual metric inspection, which was both slow and difficult. Instead we envision an automated monitoring system that collects operational metrics (e.g., each software component's information encoded as $p(x_i|pa_i^\mathcal{G})$) and detects distributional shift in an event of operational issue. This is followed by an automated change attribution calculation procedure that would allow to automatically discover the faulty service and only engage engineer responsible for operating that service, saving time and effort for Nadia and her colleagues.

\subsection{Business analysis}
Business analysts also often require tools to identify root causes of unexpected shifts in business metrics and Key Performance Indicators (KPI) \cite{ambler2000marketing, carpi2017performance}. The process of answering such questions bears some resemblance to technical troubleshooting discussed above, as business analysts often require understanding of connections between input and output data. The main difference is their concern with business metrics and KPIs \cite{cadle2010business}. Therefore analysts need to backtrack the effect of complex software computations for a given business metric and attribute its change to highlight which original input data factors contributed the most to the value of interest. 

We envision an automated tool that, given the observed output, applies Algorithm~\ref{alg:attribution_change} and returns the contribution of each input data stream. In our example 2 Art could use such tool to calculate effects of each of suspected reasons to the observed drop in sales. Relative magnitude of the calculated effects could be used to understand how certain the estimates are and whether any additional investigation is required.

\subsection{Experimentation}
It is difficult to do efficient experimentation in large software systems. Isolated test environments with generated traffic can be an efficient way to verify correctness of new changes before they are pushed to production. However such test environments have to be complete replicas of the production setting, which means engineers have to maintain two separate identical stacks. Unfortunately, in practice test stacks tend to be de-prioritised and fall behind in deployment schedule \cite{strublemodel}, which eventually leads to situations where successful deployment on the test stack does not indicate issue-free deployment to production \cite{ramler2014practical, page2008we}. A/B testing on live system is another popular approach to validate effect of new changes. However A/B tests are hard to design correctly \cite{olsson2017experimentation} and may have unforeseen side effects on the entire live system \cite{xu2015ABchallenges}. It is also an expensive practice, as it requires large amount of traffic from real user base to generate statistically valid results \cite{kohavi2022ABbusters}.

Causal inference on dataflow graphs offers an alternative solution to experimentation in software. Once developers have identified node(s) in the graph that are planned to be updated and are able to quantify output distribution, post-intervention distributions across the entire system can be calculated with do-calculus. In our example 3 Alice could retrieve historical data for demand, measure output distribution of the new demand prediction model, and estimate how the new model may affect other CoffeeFlow components.

\setlength{\textfloatsep}{\textfloatsepsave}

\section{Experiments}
\begin{figure}[t]
	\centering
	\includegraphics[width=\columnwidth]{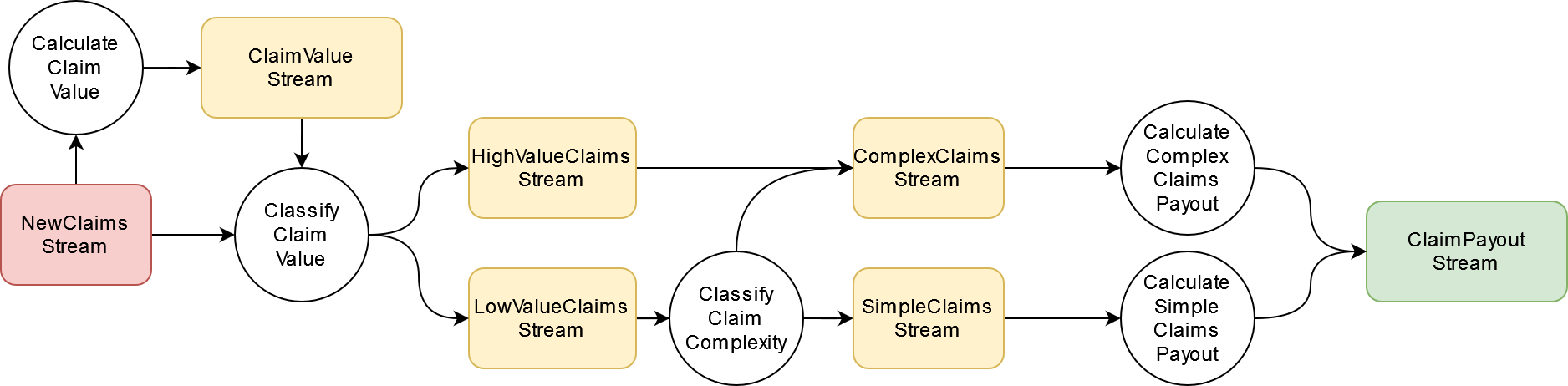}
	\caption{Dataflow diagram of the FBP application for insurance claims processing.}
	\label{figure:insurance_claims}
\end{figure}

This section illustrates our idea on an FBP program for insurance claims processing \cite{Paleyes2022FBP}. The program models processing of a car insurance claim that goes through a series of decision making steps, before the final payout amount is calculated. Its dataflow graph is shown on Figure \ref{figure:insurance_claims}.

Our first experiment\footnote{Our experiments are implemented with DoWhy-GCM library \cite{dowhy_gcm} and use independence tests and Shapley values, as discussed in Section \ref{sec:dataflow_graph_as_causal_graph}. Full code can be found at \url{https://github.com/apaleyes/dataflow-causal-graph}.} demonstrates usage of causality for fault localisation. To deploy a software bug, we break the logic of the \textit{ClassifyClaimComplexity} node, causing it to classify all low value claims as simple. That change has a clear downstream effect on the final payouts. We collect raw data from the streams before and after the bug, confirm distribution shift of the claim payouts, and use Algorithm \ref{alg:attribution_change} to calculate the attributions of each node of the graph in the observed shift.

\begin{table}
\caption{Examples of attribution scores and corresponding probabilities for a single run of fault localisation (columns 2 and 3) and data shift (columns 4 and 5) experiments. Scores for most likely sources of the output distribution shift in each experiment are highlighted in bold text.}
\begin{center}
\begin{tabular}{ |c|c|c|c|c| } 
    \hline
    & \multicolumn{2}{|c|}{Experiment 1} & \multicolumn{2}{|c|}{Experiment 2} \\
    & \multicolumn{2}{|c|}{Fault localisation} & \multicolumn{2}{|c|}{Data shift} \\
    \cline{2-5}
    Node Name & Score & Probability & Score & Probability \\
    \hline
    NewClaimsStream & 0.0011 & 0.04 & \textbf{0.1995} & \textbf{0.66} \\ 
    ClaimValueStream & 0.0033 & 0.11 & -0.0043 & 0.01 \\ 
    LowValueClaimsStream & 0.0012 & 0.04 & -0.0151 & 0.05 \\ 
    HighValueClaimsStream & 0.0009 & 0.03 & 0.0613 & 0.20 \\ 
    SimpleClaimsStream & \textbf{0.014} & \textbf{0.47} & 0.0083 & 0.03 \\ 
    ComplexClaimsStream & -0.0087 & 0.29 & 0.0066 & 0.02 \\ 
    ClaimPayoutStream & 0.0004 & 0.01 & 0.0078 & 0.03 \\ 
    \hline
\end{tabular}
\end{center}
\label{table:experimental-results}
\vspace{-15pt}
\end{table}

\begin{figure}[t]
	\centering
	\includegraphics[width=\columnwidth]{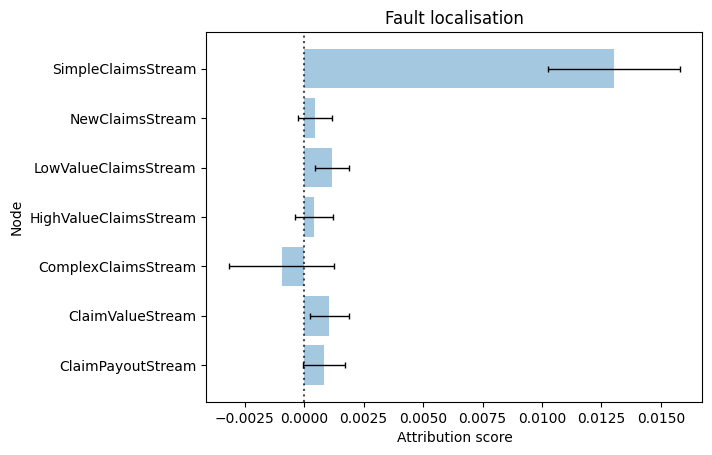}
	\caption{Mean attribution score of each data node after 30 repeats of the fault localisation experiment. Black segments indicate confidence intervals.}
	\label{figure:fault-localisation-exp-result}
\end{figure}

\begin{figure}[t]
	\centering
	\includegraphics[width=\columnwidth]{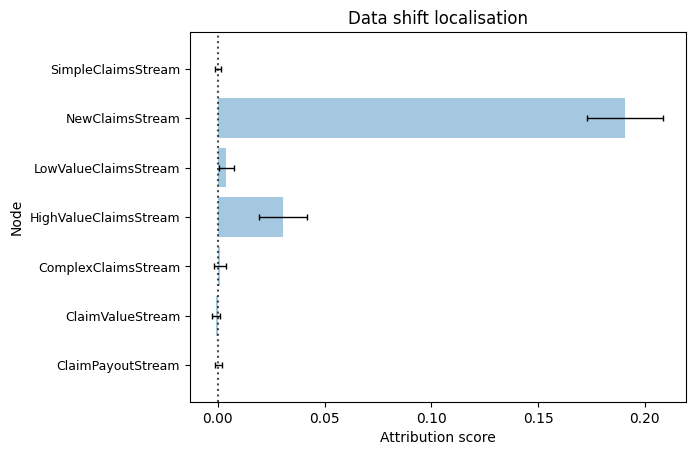}
	\caption{Mean attribution score of each data node after 30 repeats of the data shift experiment. Black segments indicate confidence intervals.}
	\label{figure:data-shift-exp-result}
\end{figure}

Table \ref{table:experimental-results} (columns 2 and 3) shows an example of the algorithm output for the fault localisation experiment. Both raw attribution values as well as the corresponding probabilities are shown. \textit{SimpleClaimsStream} is identified as the most likely source of the observed shift, and as this stream is an output of the \textit{ClassifyClaimComplexity} node, we can correctly deduce that this node is indeed the source of a problem. This table presents the information in a way an engineer might see it when similar technology is deployed in a production monitoring system. They can also use this information to understand how confident the attribution procedure is. Figure \ref{figure:fault-localisation-exp-result} shows the results of repeating this experiment 30 times on randomised input data. Welch’s t-test gives $p < 0.01$ indicating statistical robustness of our procedure in identifying the source of the bug in this experiment.

The second experiment uses causality to identify an input data shift. The setup of this experiment is very similar except this time we perturb input data stream \textit{NewClaimsStream} with increasing originally claimed amount by 50\%. We expect Algorithm \ref{alg:attribution_change} to attribute the change in the output distribution to the input data node and not to any intermediate nodes.

Table \ref{table:experimental-results} (columns 4 and 5) shows an example of the algorithm output for the data shift experiment. As expected, shifts in all intermediate data streams are ignored, and the input data stream has the highest attribution score. Again, quantitative uncertainty information is available to estimate how reliable the output is. Figure \ref{figure:data-shift-exp-result} shows the results of repeating this experiment 30 times on randomised input data. $p < 0.01$ on Welch’s t-test shows that identifying \textit{SimpleClaimsStream} as the most likely node is a statistically significant outcome.

\section{Conclusions and research agenda}
With this work we draw the attention of the community to the FBP paradigm and encourage more applications of causality to systems. We observed the connection between dataflow architectures and causal graphs, discussed its potential applications, and illustrated how that idea could work on two simple experiments. As a future work, we plan to build prototypes of automated tools described in Section~\ref{sec:dataflow_graph_as_causal_graph} to validate the idea further. We also plan to investigate how causal inference on FBP dataflow graphs can assist in producing system-wide counter-factual explanations for GDPR compliance \cite{wachter2017counterfactual}. Furthermore, we believe the synergy between dataflow architecture and causality opens an exciting research agenda for the community, and discuss a few directions below.

Engineers need a thorough understanding of the trade-offs involved in choosing FBP over other methodologies. Platforms, tools and frameworks are key considerations when starting a new software project. For example, message queues, data streams, or databases can all be utilised as data connections in the graph, each option with its pros and cons. Failure modes of dataflow pipelines constitute another area of research, as well as identification of key metrics that reflect health of a dataflow system and its components. The scalability of current causal inference techniques is still unclear. As modern software systems tend to be rather complex, corresponding dataflow graphs can get large. This aspect puts emphasis on the effort to scale modern causal inference methods to apply them to bigger and more complex graphs. Concerns over associated costs and storage requirements should also be investigated.

We hope that this work highlights an interesting opportunity to research understanding of dataflow graphs in software and their connection to causality, leading to improvements in software usage and development.




\bibliographystyle{abbrv}
\bibliography{bibliography}

\end{document}